# Technical Report # KU-EC-13-1
# Censoring Distances Based on Labeled Cortical Distance Maps in Cortical Morphometry

E. Ceyhan[1*], T. Nishino[3,4], J. Alexopolous[3], R. D. Todd[5], K. N. Botteron[3,4], M. I. Miller[2,6,7], J. T. Ratnanather[2,6,7]

November 10, 2012

[1]*Dept. of Mathematics, Koç University, 34450, Sarıyer, Istanbul, Turkey.*
[2]*Center for Imaging Science, The Johns Hopkins University, Baltimore, MD 21218.*
[3]*Dept. of Psychiatry, Washington University School of Medicine, St. Louis, MO 63110.*
[4]*Dept. of Radiology, Washington University School of Medicine, St. Louis, MO 63110.*
[5]*Dept. of Genetics, Washington University School of Medicine, St. Louis, MO 63110.*
[6]*Institute for Computational Medicine, The Johns Hopkins University, Baltimore, MD 21218.*
[7]*Dept. of Biomedical Engineering, The Johns Hopkins University, Baltimore, MD 21218.*

\*corresponding author:
Elvan Ceyhan,
Department of Mathematics, Koç University,
Rumelifeneri Yolu, 34450 Sarıyer,
Istanbul, Turkey
e-mail: elceyhan@ku.edu.tr
phone: +90 (212) 338-1845
fax: +90 (212) 338-1559

## Abstract

**Background**: It has been demonstrated that shape differences are manifested in cortical structures due to neuropsychiatric disorders. Such morphometric differences can be measured by labeled cortical distance mapping (LCDM) which characterizes the morphometry of the laminar cortical mantle of cortical structures. LCDM data consist of signed/labeled distances of gray matter (GM) voxels with respect to GM/white matter (WM) surface. Volumes, thickness and descriptive measures (such as means and variances) for each subject and the pooled distances can help determine the morphometric differences between diagnostic groups, however they do not reveal all the morphometric information contained in LCDM distances. To extract more information from LCDM data, censoring of the pooled distances is introduced for each diagnostic group. For censoring of LCDM distances, the range of LCDM distances is partitioned at a fixed increment size; and at each censoring step, the censoring distance is increased as the increment size, and distances not exceeding the censoring distance are kept.
**Results**: Censored LCDM distances inherit the advantages of the pooled distances. Furthermore, the analysis of censored distances provides information about the location of morphometric differences which cannot be obtained from the pooled distances. However, at each step, the censored distances aggregate, which might confound the results. The influence of data aggregation is investigated with an extensive Monte Carlo simulation analysis and it is demonstrated that this influence is negligible. As an illustrative example, GM of ventral medial prefrontal cortices (VMPFCs) of subjects with major depressive disorder (MDD), subjects at high risk (HR) of MDD, and healthy control (Ctrl) subjects are used. A significant reduction in laminar thickness of the VMPFC and perhaps shrinkage in MDD and HR subjects is observed when compared to Ctrl subjects with significant morphometric differences occurring at different GM LCDM distance values for MDD and HR subjects.
**Conclusions**: The methodology is also applicable to LCDM-based morphometric measures of other cortical structures affected by disease.

# 1 Background

Recent advances in high resolution magnetic resonance imaging (MRI) technology and developments in computational anatomy (CA) methods (see, e.g., [1-6]) have resulted in a substantial increase in understanding of the laminar structure of the neocortex. In neuroimaging studies of neuropsychiatric and neurological diseases such as Alzheimer's disease and schizophrenia, Labeled Cortical Distance Mapping (LCDM) [6] has been proven to be a powerful tool for quantization and comparison of cortical thickness features with specific examples demonstrated in the cingulate cortex [7-10], occipital cortex [11-12], frontal cortex [13-14] and superior temporal gyrus [15] . Though developed for cortical regions, LCDM profiles for whole brains are similar in shape [16-17]; further LCDMs have been adapted or modified to deal with deeply buried sulci [18-19].  LCDMs are comparable to other methods for computing cortical thickness [20].

LCDM data are distances of labeled gray matter (GM) voxels with respect to the gray/white matter (GM/WM) cortical surface, and so quantize and characterize the morphometry of the laminar cortical mantle. Here "morphometry" has two components, the structural formation (like surface and form of the tissue) and scale or size (like volume and surface area). Thus, morphometry refers to all aspects of laminar shape, where "shape" refers to the surface structure, and "size" refers to the scale of the tissue in question. Analysis of volumes (in $mm^3$), of descriptive measures (i.e., summary statistics) of pooled distances, and of the pooled distances yield "rough" comparisons of cortical regions of interests (ROIs) between groups, in the sense that, if significant, a comparison indicates global morphometric (shape and/or size) differences in cortical ROIs between groups [21-22]. But they do not reveal where (e.g., at which distance from GM/WM surface) these differences occur. As the LCDM distances measure the distance from GM voxel centers to GM/WM surface, they carry more than just shape/size information. This suggests that, properly used, LCDM distances may also provide at which distance GM in the cortical ROI differ between groups, thereby providing additional information about the underlying nature of the difference associated with the disease.

Abnormalities have been demonstrated in structure and function of specific regions of the prefrontal cortex associated with major depressive disorder (MDD) [23-24]. Previous structural imaging studies have largely focused on adult onset MDD, while only a few have focused on early onset MDD. Structural deficits in a subregion of the VMPFC, i.e., subgenual prefrontal cortex, have also been associated with early onset of MDD [25-30]. LCDM data for the Ventral Medial Prefrontal Cortex (VMPFC) has been  analyzed in detail [21-22]. Here, the data based on a twin design neuroimaging study contained three diagnostic groups, namely, MDD, being at high risk (HR) for MDD, and the control (Ctrl) group. In [22], morphometric summary measures such as mean, median, variance, etc. of the LCDM distances and volumes were analyzed, but these summary statistics failed to detect differences between MDD and healthy subjects. Since such measures were oversimplifying the vast amount of information in LCDM data, pooling of the LCDM distances by diagnostic group, rather than subsampling, was introduced so as to detect morphometric differences with a higher sensitivity [21]. In pooled LCDM distances, the entire LCDM data set was used, and the validity of the underlying assumptions for the tests was investigated. Significant morphometric differences in VMPFC were observed associated with MDD or being at HR for MDD.

The automated methods for measuring cortical thickness in MRI are classified as voxel-based, surface-based or a mixture of the two [16].  In our LCDM approach, the surface between GM/WM is determined, and then distance of each voxel to this surface is computed [6-7]. This approach has also been extended by modeling image intensity stochastically based on the normal distance where the model includes cortical thickness as one of the parameters [31]. In this regard, our LCDM approach is very similar to the voxel-based cortical thickness (VBCT) method of [16] where each voxel in the GM has a thickness value associated with it, but our analysis of these voxel-based thickness values is different. In [16] cortical thickness values are compared on a voxel-by-voxel basis as in SPM2 (see [32]), while in our analysis of LCDM distances, we first pool (i.e., merge) the distance values for each diagnostic group, and perform the comparisons on the overall distance (or thickness) level, rather than the voxel level for each individual. Previously, we have discussed the analysis of these pooled distances for the overall comparison of morphometric differences due to depression [21].

In this article, we propose censoring of LCDM distances which may provide more information about the distribution of GM voxels. In censoring, we partition the range of LCDM distances at a particular increment size, and at each increment, we only keep LCDM distances not exceeding the corresponding censoring distance relative to the GM/WM surface. As an illustrative example, we use the same LCDM data for VMPFC as in [21-22] so as to demonstrate the benefits of censoring LCDM distances compared to pooled LCDM distances. Censored LCDM distances inherit the advantages of the pooled distances (such as robustness to assumption violations and sensitivity to morphometric differences due to a disease) and also provide information on the laterality and location of changes associated with the disease in question. In particular, by using the censored distances, one can determine where significant differences in GM of VMPFC occur related to MDD or HR in terms of distance to the GM/WM surface. By Monte Carlo simulations, we demonstrate that comparison of censored distances between diagnostic groups is robust to the violations of the underlying assumptions such as within sample independence and normality (i.e., Gaussianity). Furthermore, at each censoring step distances less than or equal to the corresponding censoring distance aggregate, which might confound the results of the analysis. Our extensive Monte Carlo study also indicates that such an aggregation effect is negligible for censored distances. Additionally, censored distances are very sensitive to indicate differences as a function of distance from the GM/WM surface.

We describe the example data and its acquisition in Section 2.1, censoring methods in Section 2.2, statistical methodology in Section 2.3, analyze the censored distances in Section 3, and investigate the influence of aggregation of censored distances and assumption violations with an extensive Monte Carlo simulation study in Section 4.

## 2    Methods
## 2.1  Data description and acquisition

For details of the MRI tools and methodology to prepare VMPFC to measure LCDM distances, and for specifics of the measurement process of LCDMs, see [21] with respective references. Briefly, a cohort of 34 right-handed young female twin pairs (ages between 15 and 24) were obtained from the Missouri Twin Registry, to study MDD related cortical changes in the VMPFC. Both monozygotic and dizygotic twin pairs were included, of which 14 pairs were controls and 20 pairs had one twin affected with MDD, and their co-twins were designated as the HR group. To obtain LCDM distances, we partition the ROI by a regular lattice of voxels of size $h$, labeling each as GM, WM, or cerebrospinal fluid (CSF) (see, e.g., [6, 33]). For every voxel in the ROI subvolume, the distance from the voxel centroid to the closest point on GM/WM surface is measured and generally signed according to the type of the voxel, e.g., positive for GM and CSF, and negative for WM.

Reliability of LCDMs is dependent on reliability of GM segmentation and reconstruction of GM/WM surfaces which has been validated for several cortical structures including VMPFC [34], cingulate cortex [8, 35] and planum temporale [36].

Figure 1 illustrates the kernel density estimate of LCDM distances of GM voxels of a typical cortical structure of interest. In this cortical structure most of GM distances are positive. If two LCDM distance sets are different (with everything else being same), one can safely deduce that the corresponding VMPFCs have different morphometric structures. Thus, LCDM may serve as a tool to analyze and/or compare the morphometry (shape and size) of cortical tissues in brain.

Let $D^L$ ($D^R$) be the set of LCDM distances for the left (right) ROI. Then $D^L = \{D_{ijk}^L, i=1,2,3, j=1,2,\ldots,n_i, k=1,2,\ldots,K_{ij}\}$ where $D_{ijk}^L$ is the LCDM distance for the $k^{th}$ voxel in GM of left VMPFC of subject $j$ in group $i$ (with $i=1$ for MDD, $i=2$ for HR, and $i=3$ for Ctrl). So, we have $n_1 = 20$, $n_2 = 20$, and $n_3 = 28$. Right VMPFC distances $D^R$ are denoted similarly as $D^R = \{D_{ijk}^R, i=1,2,3, j=1,2,\ldots,n_i, k=1,2,\ldots,K_{ij}\}$. We only retain LCDM distances between -0.5 and 5.5 *mm* based on prior anatomical knowledge (e.g., [37]), so as to avoid voxels potentially mislabeled as GM. By doing so only a negligible portion ($\leq 0.20\%$) of the distances are removed from further analysis.

## 2.2 Censoring LCDM distances

To obtain information on the location of morphometric differences measured as distance to the GM/WM surface, we propose the following procedure, which is called *censoring of LCDM distances*. We first partition the set of LCDM distances into bins of size $\delta$, then we have $\lfloor d_{\max}/\delta \rfloor$ many bins where $d_{\max}$ is the largest LCDM distance value in $D^L \bigcup D^R$ and $\lfloor s \rfloor$ stands for the floor of $s$, i.e., largest integer less than or equal to $s$. In order to construct LCDM censored distances, we only retain distances less than or equal to the specified distance value. That is, at $k^{th}$ censoring step, we only consider the voxels whose LCDM distance is less than or equal to $\gamma_{k,\delta} = k\delta$. Thus we only consider the layer of the cortex with distance of $\gamma_{k,\delta}$ or less from the GM/WM surface. These distances are the censored LCDM distances, which, for left VMPFCs, are denoted as

$$C_d^L(k,\delta) := \{d \in D^L \cap [-0.5, k\delta]\} = \{d \in D^L : d \leq k\delta\} \tag{1}$$

and for group $i$ in left VMPFCs,

$$C_{d,i}^L(k,\delta) := \{d \in D_i^L : d \leq k\delta\} \tag{2}$$

for $i = 1,2,3$ (i.e., for groups MDD, HR, and Ctrl, respectively). Censored LCDM distances for right VMPFCs are denoted similarly as $C_d^R(k,\delta)$ and as $C_{d,i}^R(k,\delta)$ for group $i$, respectively. This procedure is called censoring, because distances are measured for voxels, if the centroids of the voxels are closer to the GM/WM surface than a threshold, and the distances for the remaining voxels are not measured. By censoring LCDM distances, we partition the VMPFCs with respect to distance from GM/WM surface; thereby can obtain more detailed and localized morphometrics of the VMPFCs. For example, if analysis of censored distances yields a significant result at step $k$, it would indicate significant morphometric differences between diagnostic groups at GM distance of $k\delta$ *mm*. If significant differences are observed at all censoring steps between $k$ and $l$, then this would mean that significant morphometric differences occur for GM distance values between $k\delta$ and $l\delta$ *mm*.

In the following sections, we use $d_{\max} = 5.5$ *mm* and we pick $\delta = 0.01$ *mm*, hence $k = 0,1,2,\ldots,550$ and $\gamma_{k,\delta} = 0.00, 0.01, 0.02,\ldots, 5.50$ *mm*. Due to the confounding influence of mislabeled GM voxels close to the GM/WM surface, censoring distances in [1,5.5] *mm* pro-vide more reliable results. Note also that for $\gamma_{k,\delta} = 5.5\,mm$, i.e., at the last censoring step, the censored distance analysis is identical to the pooled distance analysis provided in [21].

## 2.3 Statistical methodology

For a specific subject, the LCDM distances for neighboring voxels are correlated; hence there is an inherent spatial dependence between LCDM distances at the individual level. Pooling and censoring do not remove this dependence; on the contrary, they ignore the subject information but only take diagnostic group information into account. In [21], we considered Kruskal-Wallis (K-W) and ANOVA *F*-tests for multi-group comparisons and Wilcoxon rank sum test and Welch's *t*-test for pairwise comparisons of pooled distances. Furthermore, Kolmogorov-Smirnov (K-S) test was employed for distributional comparisons (see [38] for information on these tests). However these tests only detect global morphometric differences but do not provide where in the ROI (e.g., VMPFCs) these differences occur. In [21], it is demonstrated that the influence of these assumption violations is negligible.

We introduce censoring of LCDM distances to find out where (i.e., at which distance value) the significant differences occur. For left (and right) censored distance comparisons, at each censoring step, we apply K-W test for the equality of the distributions for all (three) groups and ANOVA *F*-test with and without assuming homogeneity of the variances of the distances. These tests are used to detect possible differences between groups in these censored distances. If K-W test yields a significant *p*-value at a censoring distance value, then the morphometry is different for at least two of the groups and this difference starts to occur at this censoring distance value. To find out which pairs of groups exhibit significant morphometric differences at this distance, we use pairwise Wilcoxon rank sum test to compare the pairs of the groups. Similarly, if one of the ANOVA *F*-tests is significant, then we use pairwise *t*-test to compare the pairs of the groups. We perform similar analyses for right censored LCDM distances. See, e.g., [38] for more detail on the tests.

When applied on censored LCDM distances, K-W test and Wilcoxon tests may provide at which distance the distributional differences occur, and ANOVA *F*-tests and Welch's *t*-tests might provide at which distance the mean LCDM distances start to differ. However, K-S test might be misleading when applied to censored distances, since it will indicate the distance where the first significant difference occurs, but by construction, the test will tend to yield the same (or more significant) *p*-values at subsequent censoring steps.

For each of the above tests, if the tests start to be significant at a certain censoring distance, say $d_1$ and stays significant for subsequent steps up to distance $d_2$, then the morphometric differences in the GM tissue start to be detectable by LCDM distances at voxels whose distance is at or larger than $d_1$ and the significant morphometric difference persists up to distance $d_2$. Hence the importance of the censoring of the LCDM distances: it provides not only significant morphometric differences, but also where (i.e., at which distance value) the differences are located (with respect to the GM/WM surface).

## 3 Results and discussion
## 3.1 Analysis of censored LCFM distances of VMPFCs

We have pooled the LCDM distances of subjects in the same group and kept distances between $[-0.5, 5.5]\,mm$ and at each censoring distance, $\gamma_{k,\delta}$, we have the distance values in $[-0.5, \gamma_{k,\delta}]\,mm$. These censored distances convey shape/size information at the specified $\gamma_{k,\delta}$ value, i.e., at distance of $\gamma_{k,\delta}$ or less from the GM/WM surface. At each censoring step $k$, the distribution of the censored distances (hence distribution of pooled distances) is severely non-normal based on Lilliefor's test of normality (all *p*-values are virtually zero).

Among the underlying assumptions of the parametric tests (ANOVA *F*-tests and *t*-tests), within sample independence and normality (Gaussianity) of LCDM distances are violated and for the morphometric tests (K-W and Wilcoxon rank sum tests), within sample independence is violated. However the violation of these assumptions for the pooled LCDM distances was shown to be negligible [21]. The censored LCDM distances inherit this robustness property of the pooled distances (as the censoring is performed on the pooled distances). Although more assumptions are violated for the parametric tests, we still use them, since both parametric and non-parametric tests are not influenced by these violations [21]. Furthermore, parametric tests are more sensitive against the alternatives that influence the means, while nonparametric tests are more sensitive against the alternatives that influence the ranking (i.e., ordering) of the distances. Due to the confounding effect of mislabeled voxels, we only consider the censoring distance analysis for [1.0,5.5] *mm*, as the analysis for this range will be more reliable. This cautionary measure is in effect for this sample data set, and if the problem of mislabeled voxels is minute or sufficiently small, one could consider the whole range of distances (i.e., [-0.5,5.5] *mm*).

### 3.1.1 Multi-group comparisons by censored LCDM distances

K-W test and ANOVA *F*-tests yield significant differences between LCDMs of the three groups ( $p < 0.0001$ for each multi-group test). Hence there are significant morphometric differences (in each of left and right VMPFCs) in at least two of the diagnostic groups in question. Then, we test for group differences in censored LCDM distances to see at which distance value the significant differences start to occur. The null hypothesis for K-W test for left censored distances is

$$H_o : F_1^L(k,\delta) = F_2^L(k,\delta) = F_3^L(k,\delta) \tag{3}$$

where $F_i^L(k,\delta)$ is the distribution of left censored LCDM distances at censoring step $k$ with increment size $\delta$ for group $i = 1,2,3$ (i.e., MDD, HR, and Ctrl, respectively). The null hypothesis for ANOVA *F*-test (with or without homogeneity of variances (HOV)) for left censored distances is

$$H_o : \mu_1^L(k,\delta) = \mu_2^L(k,\delta) = \mu_3^L(k,\delta) \tag{4}$$

where $\mu_i^L(k,\delta)$ is the mean of left censored LCDM distances at censoring step $k$ with increment size $\delta$ for group $i = 1,2,3$. The null hypotheses for the right censored LCDM distances are similar with $L$ being replaced with $R$.

We record the *p*-values for K-W test and ANOVA *F*-test with HOV and plot them against censoring distance (i.e., $\gamma_{k,\delta}$) values in Figure 2 where the horizontal line is located at 0.05. When the *p*-values fall below the nominal significance level of 0.05, they are deemed to be significant. Observe that the plots for K-W test and ANOVA *F*-test with HOV are very similar and so is the plot for ANOVA *F*-test without HOV (hence not presented). The alternative for K-W and ANOVA *F*-tests does not have a direction for three or more groups. So a *p*-value less than 0.05 for K-W test (ANOVA *F*-test with or without HOV) implies that for at least two groups, the distributions (means) of the distances less than or equal to $\gamma_{k,\delta}$ are different. Based on K-W test (ANOVA *F*-test with HOV); we observe that the differences between distributions (means) of left and right censored distances start to occur at about the same distance value. The distributions and means of the distances are significantly different for at least two of the groups for distance values of 2.00 *mm* or larger for left VMPFCs, and 2.20 *mm* or larger for right VMPFCs. Significant differences occur for right VMPFC distances between 0-1.2 *mm* as well, however due to confounding nature of negative VMPFC distances, this result is reliable for the range of 1-1.2 *mm*. This implies that there are significant morphometric differences due to depression at distance values of 2.00 *mm* or larger in GM of left VMPFCs and around 1-1.2 *mm* and 2.20 *mm* or larger in GM of right VMPFCs.

### 3.1.2 Pairwise comparisons by censored LCDM distances

Pairwise comparisons of the LCDM distances for the diagnostic groups indicate that LCDM distances for MDD and HR groups are not significantly different for both left and right VMPFCs ($p$-values are 0.6043 and 0.1553, respectively). The LCDM distances for both MDD and HR left and right VMPFCs are significantly smaller than those counterparts of Ctrl left and right VMPFCs ($p < 0.0001$ for all). Hence significant reduction in laminar thickness is observed in VMPFCs associated with MDD, but the same trend is also observed associated with being at high risk for MDD as well [21].

We also found at which distance values the distributions of the censored distances are different between groups. The next question of interest is which pairs of groups are different at each distance value. Along this line, we perform pairwise comparisons of censored distance values at each censoring step $k$. For left and right VMPFC distances, at each censoring distance, $\gamma_{k,\delta}$, we test for each pair of groups by Wilcoxon rank sum test for both less than and greater than alternatives, and record the corresponding *p*-values. The simultaneous hypotheses for Wilcoxon tests for left censored LCDM distances are

$$H_o : F_1^L(k,\delta) = F_2^L(k,\delta) \text{ and } F_1^L(k,\delta) = F_3^L(k,\delta) \text{ and } F_2^L(k,\delta) = F_3^L(k,\delta) \tag{5}$$

The less-than alternative for pairwise Wilcoxon tests is then

$$H_a : F_1^L(k,\delta) > F_2^L(k,\delta) \text{ and } F_1^L(k,\delta) > F_3^L(k,\delta) \text{ and } F_2^L(k,\delta) > F_3^L(k,\delta) \tag{6}$$

More precisely, $H_a$ means that MDD censored distances tend to be smaller than Ctrl censored distances and HR censored distances tend to be smaller than Ctrl censored distances and MDD censored distances tend to be smaller than HR censored distances. The greater than alternatives are similar except that the inequalities being reversed. Then we plot *p*-values against censoring distance values. We perform similar analysis for right censored distances also. The null hypotheses for pairwise *t*-tests are similar to the ones in expressions (5) and (6) with $F$ being replaced by $\mu$ and the inequalities reversed.

The *p*-values for left VMPFC groups are plotted in Figure 3, where the plots are for "MDD < Ctrl", "HR < Ctrl", and "MDD < HR" alternatives. Since the one-sided tests are complementary, in the sense that, the resulting *p*-values for the left-sided and right-sided alternatives should add up to 1, we only present the "less-than (<)" alternatives for the pairwise tests. Notice that at each plot, 0.05 and 0.95 are indicated by horizontal lines, and if the p-value falls below 0.05 (above 0.95), then the test is significant for the "less-than (<)" ("greater-than (>)") alternative. Based on the plots of the *p*-values of the "less-than (<)" alternatives for left VMPFCs, we see that MDD left censored distances tend to be significantly smaller than Ctrl left censored distances of 1.6 *mm* or higher. That is, at distance values of 1.6 *mm* or larger from the GM/WM surface, there are fewer GM voxels in MDD left VMPFCs than those in Ctrl left VMPFCs. Similarly, at distance values of 2.8 *mm* or larger from the GM/WM surface, there are fewer GM voxels in HR left VMPFCs than those in Ctrl left VMPFCs. On the other hand, MDD left censored distances are significantly smaller than HR left censored distances for $\gamma_d(k,\delta)$ values between 1.8 and 4.6 *mm*. Hence, there are fewer GM voxels in MDD left VMPFCs at distance values in [1.8, 4.6] *mm*. Based on the results of the *t*-tests (see Figure 3), we notice virtually the same results, except that mean distance for MDD left VMPFCs is significantly smaller than that of HR left VMPFCs at distances between 1.8 and 4.2, while MDD left distances tend to be smaller (in ranking) than HR distances for distances between 1.8 and 4.6 *mm*.

The *p*-values for pairwise Wilcoxon tests for right VMPFC groups are plotted in Figure 4 (plots for pairwise *t*-tests are virtually same, hence omitted). Notice that there are fewer GM voxels in MDD right VMPFCs at distance values between 0-1.5 *mm* (of which only the range 1-1.5 mm is reliable) and at 2.1 *mm* or higher compared to Ctrl right VMPFCs. Similarly, there are fewer GM voxels in HR right VMPFCs at distance values between 0-1.5 *mm* (of which only the range 1-1.5 mm is reliable) and at 2.2 *mm* or higher compared to Ctrl right VMPFCs. On the other hand, the distances for MDD and HR right VMPFCs are not significantly different for the entire range of 0-5.5 *mm*, except that MDD distances are significantly smaller for distance values around 2.2 and 2.5 *mm*.

*Remark 3.1:* **Analysis of censored LCDM distances versus distribution comparisons:** Observe that analysis of censored LCDM distances provides much more information compared to comparisons of the distribution of pooled distances. In particular, Wilcoxon test and K-S tests do not provide the distance values at which the differences occur. K-S test together with the ecdf plot might provide further details on the morphometry of VMPFCs compared to Wilcoxon test. However, ecdf plots suffer from the cumulative nature of the distances. On the other hand, kernel density estimates provide information on how frequent the voxels are at particular distance values. We present the kernel density plots of the LCDM distances for left and right VMPFCs by group in Figure 5 which suggests that smaller distances are more frequent (with respect to the total number of GM voxels for both groups) for MDD and HR VMPFCs. Furthermore, these density plots suggest that MDD and HR distances are more similar (up to, maybe, a scale factor) compared to the Ctrl distances. Observe also that the kernel density estimates agree with the results of the censored distance results plotted in Figure 3 and Figure 4. However, although kernel density plots and censored LCDM distance analysis provide similar information, we cannot assign statistical significance to the differences by just using the kernel density estimates. ∎

*Remark 3.2:* **Effect of the bin size on censored LCDM distances:** The censored distances depend on the bin size, $\delta$. The bin sizes larger than the voxel resolution $h$ will potentially oversmooth the local differences, and smaller than the decimal precision of the distances will only increase the computational burden. So, in general, we recommend bin size, $\delta$, to be at the precision of the LCDM distances up to at most the voxel resolution, $h$. That is, we suggest the use of bin size between 0.01 to 0.5 *mm* here, since if too large, censored distances do not provide the desired resolution in the distances from the GM/WM surface; and if too small, censored distances do not improve on the results of 0.01 *mm*. Hence the lower limit on the bin size is only due to practical concerns. ∎

*Remark 3.3:* **Holm's correction for simultaneous pairwise comparisons:** At each censoring distance step, we could also perform an adjustment to the *p*-values obtained from Wilcoxon rank sum tests (or *t*-tests) for simultaneous pairwise comparisons by Holm's correction method [39]. However, although we perform such a correction for the analysis of pooled distances [21], we avoid it for censored distances, since it is the consecutive list of distances that a significance persists that is more important, rather than the simultaneous compari-sons. Furthermore, after the Holm's correction is applied for simultaneous multiple comparisons for each alternative, the resulting p-values are modified, hence not complementary any more. Hence we would need separate plots for the "less than" and "greater than" alternatives. ∎

## 3.2 The influence of aggregation of censored distances and assumption violations on the tests: A Monte Carlo study

We have investigated the influence of the assumption violations due to the spatial correlation and non-normality (i.e., non-Gaussianity) inherent in the pooled LCDM distances on the tests and demonstrated that influence of such violations on the tests considered is negligible [21]. In this section we demonstrate that censored distances inherit this robustness --- to assumption violations --- of the pooled distances as well, since the censoring procedure is applied on the pooled distances. When censoring LCDM distances, at each step, the distances accumulate, which might confound the tests and their sensitivity to detect the differences between the groups. Furthermore, at each censoring step, the dependence of LCDM distances at individual level persists, and distances are significantly non-Gaussian. Here we investigate the confounding influence of such accumulation and assumption violations by Monte Carlo simulations. The most crucial step in the Monte Carlo simulation is being able to generate distances resembling LCDM distances of GM tissue in VMPFCs; i.e., simulating the true randomness in LCDM distances. For completeness, we provide the distance generation procedure, which was also described in [21].

We choose the left VMPFC of HR subject 1 (called reference VMPFC henceforth) for illustrative purposes. The LCDM distances for the reference VMPFC are denoted as $D_{21}^L$. We partition the range of distances into intervals $I_0 := [-1, 0.5]\, mm$, $I_1 := (0.5, 1.0]\, mm$, $I_2 := (1.0, 1.5]\, mm$, ... and $I_{11} := (5.5, 6.0]\, mm$. Let $v_i$ be the number of distances within inteval $I_i$, i.e., $v_i = |D_{21}^L \cap I_i|$, for $i = 0, 1, 2, ..., 11$. Then for $D_{21}^L$ we have $\vec{v} = (v_0, v_1, ..., v_{11}) =$ (2059, 1898, 1764, 1670, 1492, 1268, 814, 417, 142, 81, 61, 16). A possible Monte Carlo simulation for these distances can be performed as follows. We generate $n = 10000$ numbers with replacement in $\{0, 1, 2, ..., 11\}$ proportional to the above frequencies, $v_i$ (the choice of $n = 10000$ is because the number of distances for left VMPFC of HR subject 1 is 11659). Then we generate as many $\mathcal{U}(0,1)$ numbers for each $i \in \{0, 1, 2, ..., 11\}$ as $i$ occurs in the generated sample of 10000 numbers, and add these uniform numbers to $i$. Then we divide each distance by 2 to match the range of generated distances with $[0, 6.0]\, mm$ which is roughly the range of $D_{21}^L$. More specifically, we independently generate $n$ numbers from $\{0, 1, 2, ..., 11\}$ with the discrete probability mass function $P_0(N_j = i) = v_i/11659$ for $i = 1, 2, ..., 11$ and $j = 1, 2, ..., n$. So, $P_0(N_j = i) = v_{p,i}$ where $(v_{p,0}, v_{p,1}, ..., v_{p,11}) = \vec{v}_p =$ (0.177, 0.163, 0.151, 0.143, 0.126, 0.109, 0.070, 0.036, 0.012, 0.007, 0.005, 0.001). Let $n_i$ be the frequency of $i$ among the $n$ generated numbers from $\{0, 1, 2, ..., 11\}$ with distribution $P_0$, for $i = 1, 2, ..., 11$. Hence $n = \sum_{i=0}^{11} n_i$. Then the set of simulated distances is

$$D_{mc} = \left\{ d_{ik} = (J_i + U_i)/2 : J_i \stackrel{iid}{\sim} P_0 \text{ and } U_i \stackrel{iid}{\sim} \mathcal{U}(0,1) \text{ and } J_i \text{ and } U_i \text{ are independent for } i = 0, 1, 2, ..., n \right\}. \quad (7)$$

The Monte Carlo scheme described above generates distances that resemble LCDM distances for the reference VMPFC. Therefore, the distances generated in this fashion together with modification of some parameters such as $v_{p,i}$ would resemble the distances of VMPFCs from real subjects [21].

### 3.2.1 Simulation of realistic LCDM distances

We generate three samples $\mathcal{X}$, $\mathcal{Y}$, and $\mathcal{Z}$ each of size $n_x$, $n_y$, and $n_z$, respectively in our Monte Carlo simulations. Each sample is generated similar to the procedure described above with $n_x = n_y = n_z = 10000$. For example, we generate sample $\mathcal{X}$ as follows. Let $\eta_x$ be a positive integer less than the maximum number of voxels in the stacks in (7), namely 2059 and $v_x = (v_0^x, v_1^x, \ldots, v_{12}^x)$ with $v_i^x$ being the $i^{th}$ entry in $v_x$ such that $v_i^x$ is the $i^{th}$ value after the values $|v_i - \eta_x|$ are sorted in descending order for $i = 0,1,2,\ldots,11$ and $v_{12}^x = 11659 - \sum_{i=0}^{11} |v_i - \eta_x|$. Then we generate

$$N_{\mathcal{X}} = \{J \sim P_X, J = 1, \ldots, n_x\}, \tag{8}$$

where $P_X(J=i) = v_i^x / \sum_{i=0}^{12} v_i^x$. Furthermore, let $n_i^x$ be the frequency of $i$ among the $n_x$ generated numbers from $P_X$. Then we generate $U_{ik} \sim \text{Unif}(0, r_x)$ for $k = 1, \ldots, n_i^x$ for each $i$, where $r_x$ is a positive real number less than 2. Equivalently, the set of simulated distances for set $\mathcal{X}$ is

$$D_{mc}^{\mathcal{X}} = \left\{ (J_i + U_i)/2 : J_i \stackrel{iid}{\sim} P_0 \text{ and } U_i \stackrel{iid}{\sim} \mathcal{U}(0,1) \text{ and } J_i \text{ and } U_i \text{ are independent for } i = 0,1,2,\ldots,n_x \right\}. \tag{9}$$

Notice that the parameters that determine the set of distances are $\eta_x$ and $r_x$ with $\eta_x = 0$ and $r_x = 1$, we have distances similar to our initial choice of the reference VMPFC. Moreover, as $\eta_x$ gets larger, the distances tend to have larger values compared to the reference VMPFC, and as $r_x$ gets larger the distances tend to have more different rankings and accumulation around $k(r_x - 1)$ for $k = 1, 2, \ldots, 11$. We generate samples $\mathcal{Y}$ and $\mathcal{Z}$ in a similar fashion with parameters $\eta_y, r_y$ and $\eta_z, r_z$, respectively.

### 3.2.2 Empirical size curves

The null hypothesis, $H_o$, of multi-sample case is the equality of the distributions of LCDM distances. So under $H_o$, we generate three samples $\mathcal{X}$, $\mathcal{Y}$, and $\mathcal{Z}$ with the below parameters:

$$H_o : r_x = r_y = r_z = 1.0 \text{ and } \eta_x = \eta_y = \eta_z = 0 \tag{10}$$

Observe that each sample of $\mathcal{X}$, $\mathcal{Y}$, and $\mathcal{Z}$ is generated so as to resemble the reference VMPFC. The choice of the reference VMPFC is done without loss of generality, since any other VMPFC can either be obtained by a rescaling the distances and/or modifying the parameters. So for example, for sample $\mathcal{X}$, $P_X(X_j = i) = v_{p,i}$ with $v_{p,i}$ being the $i^{th}$ entry in $v_p$ in Section 3.2. and the set of simulated distances for set $\mathcal{X}$ is as in (9) with $r_x = 1.0$ and $\eta_x = 0$.

The censoring of the $\mathcal{X}$ distances is applied as in Section 2.1 and the censored distances are denoted as

$$C_d^{\mathcal{X}}(k, \delta) := \{d \in D_{mc}^{\mathcal{X}} \cap [0, k\delta]\} = \{d \in D_{mc}^{\mathcal{X}} : d \leq k\delta\}. \tag{11}$$

Samples $\mathcal{Y}$ and $\mathcal{Z}$ are generated similarly with generated distances are denoted as $D_{mc}^{\mathcal{Y}}$ and $D_{mc}^{\mathcal{Z}}$ and censored distances are denoted as $C_d^{\mathcal{Y}}(k, \delta)$ and $C_d^{\mathcal{Z}}(k, \delta)$, respectively.

The above data generation procedure is repeated $N_{mc} = 1000$ times. At each censoring step, we record the *p*-values for K-W test, and ANOVA *F*-tests (with and without HOV), and pairwise Wilcoxon rank sum test and *t*-test. We also count the number of times the null hypothesis is rejected at $\alpha = 0.05$ level for these tests, thus obtain the empirical significance levels (i.e., sizes) under $H_o$ in expression (10). The average *p*-values and empirical size estimates together with 95% confidence bands for K-W test are plotted against the censoring distance values in Figure 6; for pairwise Wilcoxon rank sum test for the one-sided alternatives $X < Z$ are plotted in Figure 7 (the plots for $X > Z$ are similar, hence omitted). In the left plot in Figure 6, we only plot the horizontal line at 0.05 only, since the alternative hypothesis for K-W is not one-sided. That is, there are differences between the groups for small *p*-values (which are deemed significant if smaller than 0.05). The alternative hypothesis for Wilcoxon test can be one-sided, so, if the p-values are smaller than 0.05, then sample $\mathcal{X}$ tends to be smaller than sample $\mathcal{Z}$, while if they are larger than 0.95, then sample $\mathcal{X}$ tends to be larger than sample $\mathcal{Z}$. Observe that average *p*-values are about 0.50 and empirical sizes are about 0.05 for both tests. This implies that under the null case, as expected, the simulated distances do not reveal significant differences. The empirical sizes are about the specified nominal level of .05 (i.e., the test is neither conservative nor liberal in rejecting the null hypothesis). Hence, the proposed procedure generates LCDM distance sets that not only resemble the VMPFCs of the subjects, but also possess the desired randomness in distances. That is, if the morphometry of the VMPFCs (quantified by the LCDM approach) had the same distribution for a set of subjects, their LCDM distances could have looked like the generated distances. The plots for ANOVA with and without HOV, for one-sided alternatives with pairwise Wilcoxon test for pairs $X, Y$ and $Y, Z$, and pairwise *t*-test for all three pairs are similar (hence not presented).

### 3.2.3 Empirical power curves

We consider the alternative hypotheses in which we generate sample $\mathcal{X}$ as in the null case, so $D_{mc}^{\mathcal{X}}$ is as in Equation (9). For sample $\mathcal{Y}$, we set $r_y = 1.2$ and $\eta_y = 0$ and for sample $\mathcal{Z}$ we set $r_z = 1.0$ and $\eta_z = 50$. So the alternative hypothesis we consider is

$$H_a : r_x = r_z = 1.0, r_y = 1.2 \text{ and } \eta_x = \eta_y = 0, \eta_z = 50 \qquad (12)$$

and we set $n_x = n_y = n_z = 10000$. So, $P_Y(J = i) = v_{p,i}^y$ where $(v_{p,0}^y, v_{p,1}^y, \ldots, v_{p,11}^y) = \boldsymbol{v}_p^y$ are as in Section 3.2; and $P_Z(J = i) = v_{p,i}^z$ where

$$(v_{p,0}^z, v_{p,1}^z, \ldots, v_{p,11}^z) = \boldsymbol{v}_p^z = (0.171, 0.158, 0.146, 0.138, 0.121, 0.104, 0.065, 0.051, \qquad (13)$$
$$0.031, 0.008, 0.003, 0.003, 0.001).$$

Notice that by construction sample $\mathcal{Y}$ is generated so that the rankings of distances are more different than those of sample $\mathcal{X}$ rather than the distances from the GM/WM surface. Furthermore, sample $\mathcal{Y}$ contains distances that are more accumulated at intervals [0.5,0.6], [1.0,1.1], …,[5.5,5.6] compared to sample $\mathcal{X}$. Therefore, at distances around these intervals (i.e., around $\gamma_{k,0.01}$ for $k = 50, 100, \ldots, 550$ or around $\gamma_{k,0.01} = 0.5, 1.0, \ldots, 5.5$), the censored distances for sample $\mathcal{X}$ tend to be smaller than censored distances for sample $\mathcal{Y}$. On the other hand, comparing $\boldsymbol{v}_p^z$ in Equation (13) with $\boldsymbol{v}_p$ in Section 3.2, we see that sample $\mathcal{X}$ is more likely to have distances less than 4.0 compared to those of sample $\mathcal{Z}$. Hence, we expect that the censored distances for sample $\mathcal{X}$ to be smaller than censored distances for sample $\mathcal{Z}$ at $\gamma_{k,0.01}$ for $k \geq 400$ (i.e., $\gamma_{k,0.01} \geq 4.0$). Likewise, we expect that for distances larger than 4.0, the censored distances for sample $\mathcal{Y}$ to be smaller than censored distances for sample $\mathcal{Z}$. See Figure 8 for the kernel density estimates of sample distances under the alternative hypothesis in expression (12), which agrees with the above discussion.

We repeat this sample generation procedure $N_{mc} = 1000$ times and estimate empirical power by counting the number of times the null hypothesis is rejected at $\alpha = 0.05$. The average $p$-values and empirical power estimates together with 95% confidence bands versus censoring distance values for multi-group K-W test are plotted in Figure 9. Observe that there are significant differences between groups around $\gamma_{k,0.01} = 0.5, 1.0, \ldots, 3.5$, and for distances larger than 4.0 as expected. The significant differences at steps of 0.5 increments is mostly because of sample $\mathcal{Y}$, and for distances larger than 4.0 is mostly because of sample $\mathcal{Z}$. The plots for ANOVA with or without HOV are similar (hence not presented).

The average $p$-values and empirical power estimates together with 95% confidence bands versus censoring distance values for pairwise Wilcoxon rank sum tests for the left-sided alternatives $X < Y$, $X < Z$, and $Y < Z$ are plotted in Figure 10. Based on pairwise Wilcoxon test for $X < Y$ alternative, we observe that censored distances for sample $\mathcal{X}$ tend to be smaller than censored distances for sample $\mathcal{Y}$ around $\gamma_{k,0.01}$ for $k = 50, 100, \ldots, 350$ and $k \geq 350$ (i.e., around $\gamma_{k,0.01} = 0.5, 1.0, \ldots, 3.5$ and at $\gamma_{k,0.01} \geq 3.5$). For censored distances larger than 4.0, the proportions are not large enough for samples $\mathcal{X}$ and $\mathcal{Y}$ to balance the accumulation of $\mathcal{Y}$ distances around 4.0, 4.5, 5.0, and 5.5. Hence, censored distances for sample $\mathcal{Y}$ are significantly larger than those of sample $\mathcal{X}$ for $\gamma_{k,0.01} \geq 3.5$. Based on pairwise Wilcoxon test for $X < Z$ alternative, we observe that censored distances for sample $\mathcal{X}$ tend to be smaller than censored distances for sample $\mathcal{Z}$ at $\gamma_{k,0.01}$ for $k \geq 400$ (i.e., at $\gamma_{k,0.01} \geq 4.0$). For censored distances larger than 4.0, the proportions have larger weights for sample $\mathcal{Z}$. Hence, censored distances for sample $\mathcal{Z}$ are significantly larger than those of sample $\mathcal{X}$. Based on pairwise Wilcoxon test for $Z < Y$ alternative, we observe that censored distances for sample $\mathcal{Y}$ tend to be larger than censored distances for sample $\mathcal{Z}$ around $\gamma_{k,0.01}$ for $k = 50, 100, \ldots, 350$ (i.e., around $\gamma_{k,0.01} = 0.5, 1.0, \ldots, 3.5$). For censored distances larger than 4.0, the proportions are not large enough for sample $\mathcal{Z}$ to make its censored distances larger than those of sample $\mathcal{Y}$. Hence, censored distances for sample $\mathcal{Z}$ are not significantly different from those of sample $\mathcal{Y}$ for $\gamma_{k,0.01} \geq 4.0$ with virtually zero power. This also occurs because of the proportions having larger weights for distances less than 4.0, and any parameter affecting these distances have more influence in censored distance analysis. The results of pairwise $t$-tests are similar (hence not presented).

## 4 Conclusions

In this article, we introduce the censoring of the (pooled) Labeled Cortical Distance Mapping (LCDM) distances. An LCDM data set provides information on the laminar thickness of the cortical tissue. The descriptive measures such as mean, median, or variance of distances for each subject could be recorded and analyzed. However such a crude summarization of LCDM distances for each subject is associated with a loss of information conveyed by the LCDM distance. Instead of recording only limited summary statistic for each subject, we want to characterize all of the LCDM distances. To obtain an overall VMPFC for each diagnostic condition, we pool (i.e., merge) LCDM distances of the subjects in the same group (or condition) in one data set [21].

Pooled LCDM distances, when used as a single variable, provide a method to analyze heterogeneous forms of morphometric differences [21]. When the LCDM distances of the subjects in the same diagnostic group are pooled, the most common morphometric traits of the VMPFCs for that group are more emphasized. On the other hand, the morphometric traits not common for most subjects in a group but specific to a particular subject are downplayed. The most common morphometric traits in VMPFCs in a particular diagnostic group are more likely to be related to the diagnosis of the group and pooled LCDM distances carry on the most common characteristics, so they can be very sensitive in detecting the diagnosis-specific traits of VMPFCs. As a result, LCDM distances can be suggestive of the changes in VMPFC due to a disease. When pooled distances yield significant results, it implies that VMPFCs significantly differ in morphometry (shape or size) associated with the diagnostic conditions. However, it is not suggestive of the locations of such differences, which might be important for understanding the underlying neurobiology. Hence, we propose the censoring of the pooled LCDM distances in this article to further characterize the nature of the regional differences in the specified location (i.e., distance with respect to the GM/WM surface) of morphometric differences in GM due to various conditions or associated with specific diseases. When the pooled LCDM distances are censored, the locations of the most common characteristics of VMPFCs in each group are more emphasized; hence one can detect the location of the changes in VMPFC of a subject due to, say, depression. So, censored distances inherit the nice properties of the pooled distances such as the sensitivity of the pooled distances to disease specific morphometric differences. When significant results are obtained from the censored distance analysis, it provides the distance from the GM/WM surface at which cortical mantle starts to differ in morphometry. Hence compared to pooled distances, analysis of censored distances is potentially more useful for diagnostic or clinical purposes and may provide more sensitive characterization for longitudinal treatment evaluation.

We use Kruskal-Wallis (K-W) and ANOVA (with or without HOV) $F$-tests for multi-group comparisons; and Wilcoxon rank sum, and $t$-tests tests for two-group comparisons (the first of these are used to test distributional differences and the second is used to test mean differences due to a location parameter). But, all of these tests require within sample independence, which is violated due to the spatial correlation between LCDM distances of nearby voxels. However, the influence of this violation is mild or negligible for pooled distances [21]. We demonstrate that analysis of censored distances is robust to such assumption violations, by extensive Monte Carlo simulations and this is another nice property (namely, robustness to assumption violations) inherited by censored distances. Furthermore, the aggregation of censored distances for larger censoring distances is mild to negligible. Hence we recommend both parametric and non-parametric tests for censored LCDM distances, since they are more sensitive against different alternatives. In particular, K-W and Wilcoxon tests are more sensitive to distributional differences of GM voxels at about the same distance, while ANOVA $F$-tests and $t$-tests are more sensitive against the differences in the means, that is, differences in average GM distance. One caution about censoring distances is that, major significant differences for smaller distances might confound the differences for larger distance values. However, this might be overcome by using tests on the censoring distances and Kolmogorov-Smirnov (K-S) test together with empirical cdf plots.

As an illustrative example, we use GM tissue in Ventral Medial Prefrontal Cortices (VMPFCs) as the ROI for three groups of subjects; namely, subjects with major depressive disorder (MDD), subjects at high risk (HR) for MDD, and healthy control subjects (Ctrl). Our study comprises of (MDD, HR) and (Ctrl, Ctrl) co-twin pairs. We found that there are significant morphometric differences between the groups at distances from the GM/WM surface of 2.00 mm or larger in the left VMPFC and between 1.00-1.20 mm and at 2.20 mm or larger in the right VMPFC. Furthermore, we see that left VMPFCs in MDD and left VMPFCs in Ctrl subjects show significant morphometric differences at distances of 1.60 mm or larger with significant reduction in left VMPFC associated with a history of major depression. Similarly, left VMPFCs in HR and Ctrl subjects are significantly different at distance values of 2.80 mm or larger with significant reduction in the left VMPFC in HR. That is, left VMPFC of MDD subjects tend to shrink more, since significant morphometric differences start to occur at 1.60 mm in MDD and 2.80 mm in HR left VMPFCs. On the other hand, left VMPFC in MDD is significantly smaller than HR at distances between 1.80 and 4.60 mm. Right VMPFCs in MDD and Ctrl subjects are significantly different at distances between 1.00-1.50 mm and at 2.10 mm or higher, with significant reduction in MDD. Similarly, right VMPFCs in HR and Ctrl are significantly different for distances between 1.00-1.50 mm and at 2.20 mm or higher. That is, in terms of distances, MDD and HR right VMPFCs tend to shrink but slightly more for MDD right VMPFCs (distance values of 2.20 mm for HR and 2.10 mm for MDD and between 1.00-1.50 mm for both) compared to Ctrl right VMPFCs. Right VMPFCs in MDD and HR are not significantly different for distances except around 2.20 and 2.50 mm. We thus observe a significant reduction in laminar thickness of the VMPFC and perhaps shrinkage in MDD when compared to Ctrl subjects. A similar trend can also be observed when HR is compared with the Ctrl LCDM distances. But significant morphometric differences occur at different GM distance values. These findings suggest that differences in the right VMPFC are not a consequence of episodes of MDD, but these differences are associated with higher genetic risk of MDD. Therefore, censored distances provide much more detailed information compared to pooled distances, and more powerful to help identify the local implications of the disease in the ROI.

At the microscopic level, the cortical mantle is thought to be composed of six cortical layers that are numbered I to VI as one goes from the outer or pial, i.e., GM/CSF boundary away from the skull inwards to the GM/WM surface [40]. Each layer is thought to comprise of different cells such as neuronal, pyramidal, nonpyramidal and glial cells that are important in neurotransmission between the different layers as well as with other cortical and subcortical regions [40]. Estimates of neuronal and glial densities in different cortical regions have been obtained from several histopathological, i.e., postmortem studies in humans and mammals. Reduced measures have been suggested as plausible explanations for cortical thinning observed in several neuroimaging studies albeit at the macroscopic level (e.g., [41]). In particular, in a histopathological study of major depression in humans, [42] showed both reduction in neuronal and glial density in subregions of the prefrontal cortex but that reduction in glial density was unique in the dorsolateral prefrontal cortex. Specifically they showed differences in densities in the upper cortical layers (II-IV) i.e., at distances far from the GM/WM surface or equivalently close to the pial surface in the rostral orbitofrontal regions and in the lower cortical layers (V-VI) i.e., at distances close to the GM/WM surface in the caudal orbitofrontal regions. Differences have also been demonstrated in a subportion of the VMPFC, the subgenual prefrontal cortex [42-43]. While no histopathological studies of the overall VMPFC have been done, it is conceivable that differences in censored LCDMs at distances from the GM/WM surface may be characterized by corresponding density changes. However, no definitive conclusion can be reached until LCDM analysis of a specific cortical region can be correlated with histopathological measures in an animal model of a neuropsychiatric disorder.

In summary, we have shown how LCDM distances can be used to estimate the location of differences in the cortical mantle (in terms of distance from the GM/WM surface), if censoring is performed after pooling. Such an approach can be used to analyze other cortical structures implicated in various neuropsychiatric and neuro-developmental disorders.

# List of abbreviations (in order of appearance)
VMPFC: ventral medial prefrontal cortex
LCDM: labeled cortical distance mapping

GM: gray matter
WM: white matter
MDD: major depressive disorder
HR: high risk
Ctrl: healthy control
MRI: magnetic resonance imaging
CA: computational anatomy
ROI: region of interest
VBCT: voxel-based cortical thickness
CSF: cerebrospinal fluid
K-W: Kruskal-Wallis
K-S: Kolmogorov-Smirnov
HOV: homogeneity of variances

# Acknowledgments

Research supported by R01-MH62626-01, P41-EB015909, R01-MH57180.

# Illustrations and figures

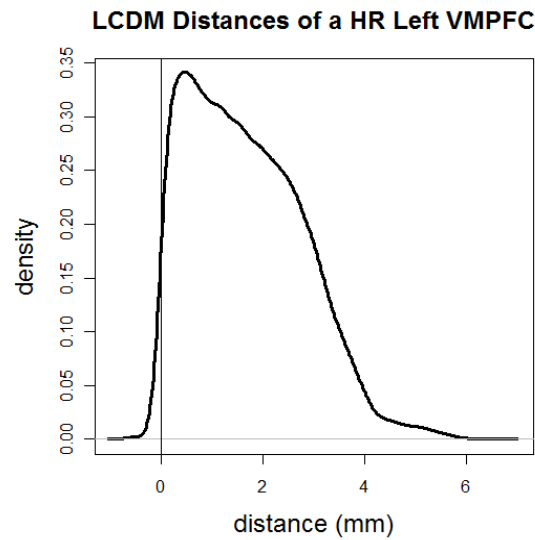

**Figure 1:** Kernel density estimate of directed (i.e., signed) LCDM distances of GM voxels for a sample cortical structure of interest. More specifically distances are for the GM of the left VMPFC of a HR subject.

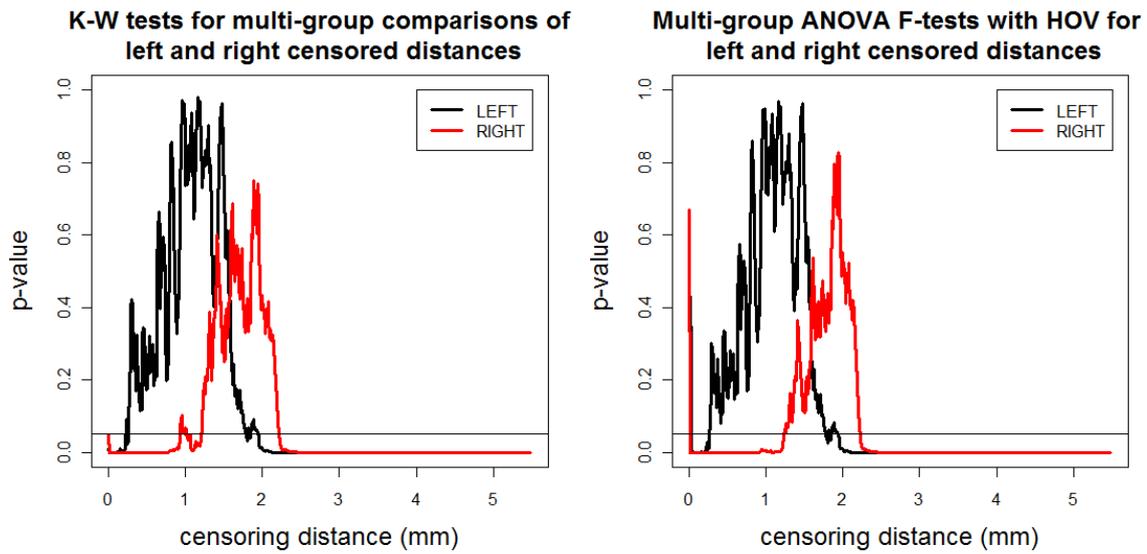

**Figure 2:** The *p*-values versus censoring distance values (*mm*) for multi-group K-W test (left) and ANOVA *F*-test with HOV (right) for LCDM distances for left (thick black solid line) and right (thin red solid line) VMPFCs. Horizontal lines are located at .05 to indicate the threshold values for significance.

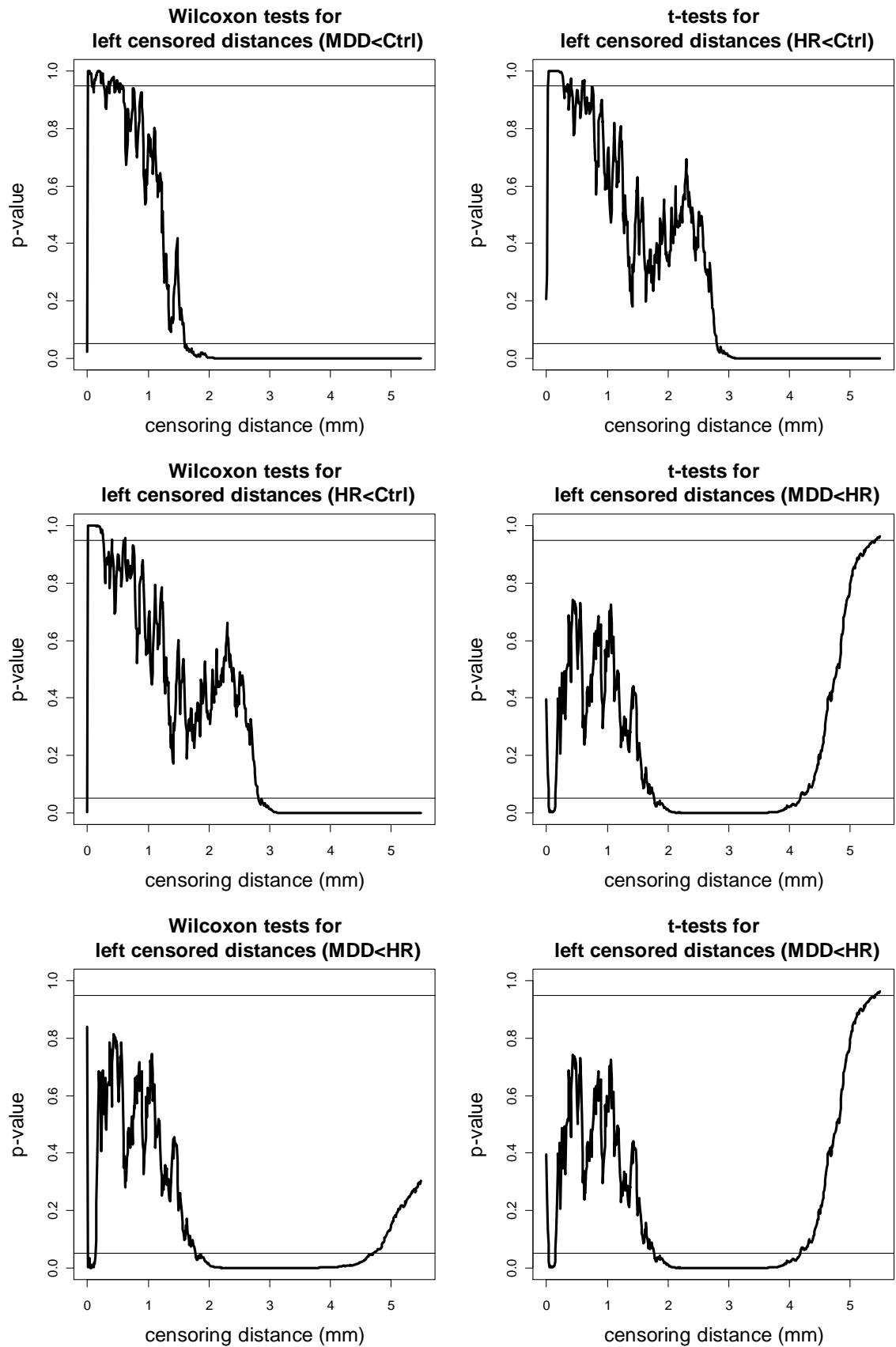

**Figure 3:** The *p*-values versus censoring distance values (*mm*) for pairwise comparisons of left VMPFC distances with Wilcoxon rank sum test (left) and t-test (right) for the less than alterna-

tive. Horizontal lines are located at 0.05 and 0.95. <: the alternative for "first less than second"; >: the alternative for "first greater than second".

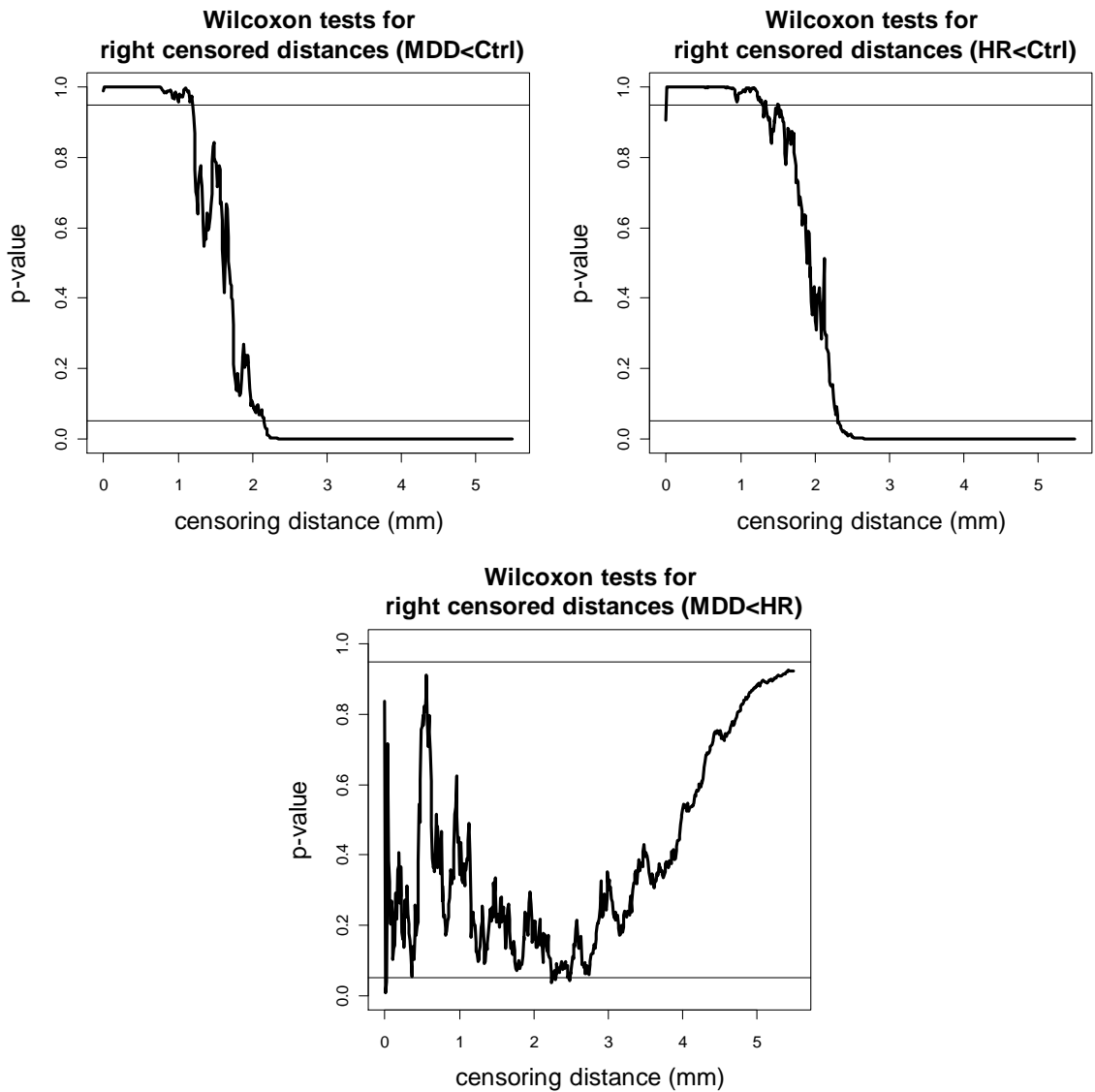

**Figure 4:** The *p*-values versus censoring distance values (*mm*) for pairwise comparisons of right VMPFC distances with Wilcoxon rank sum test for the less than alternative. Horizontal lines and alternatives are as in Figure 3.

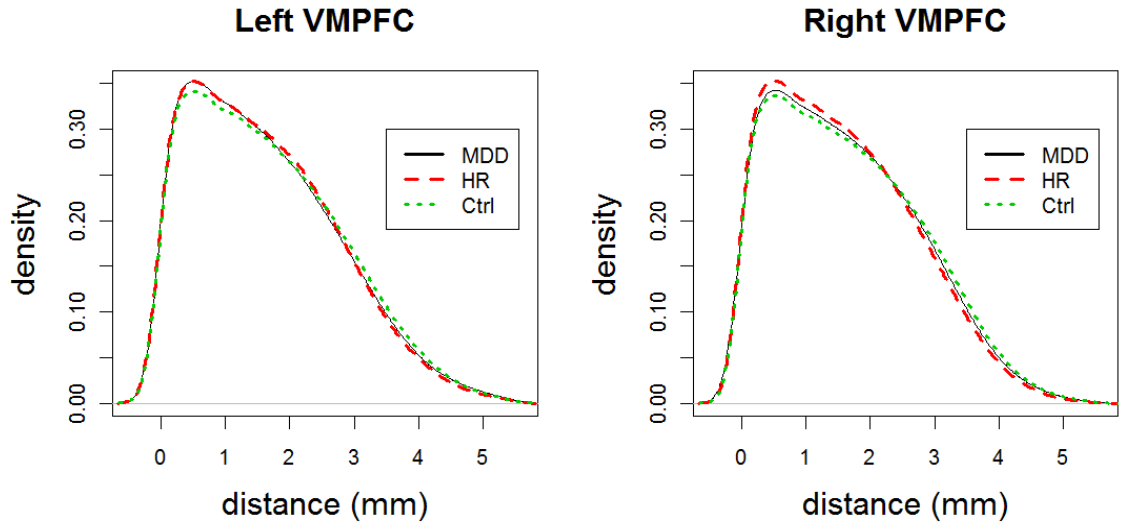

**Figure 5:** The plots for the kernel density estimates of the pooled LCDM distances for each of MDD, HR, and Ctrl groups for left and right VMPFCs.

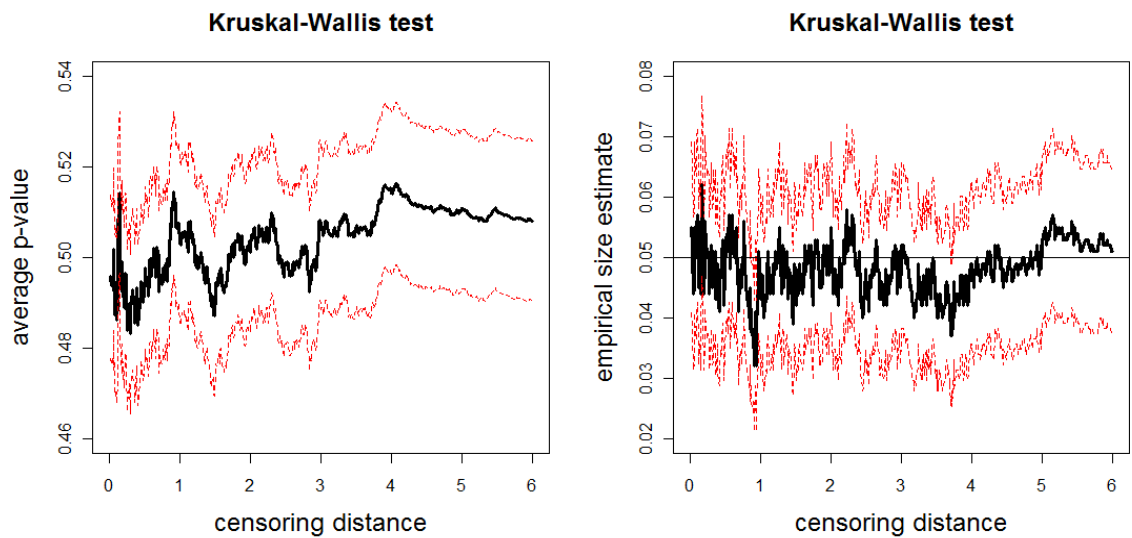

**Figure 6:** Plotted in the left is the average *p*-values (solid line) and in the right is the empirical size estimates (solid line) versus censoring distance values for multi-group K-W test together with the 95% confidence bands (dashed lines) based on 1000 Monte Carlo replications under the null hypothesis in expression (10) which implies distributional equality of censored $\mathcal{X}$, $\mathcal{Y}$, and $\mathcal{Z}$ values. Horizontal line in the right plot is at 0.05.

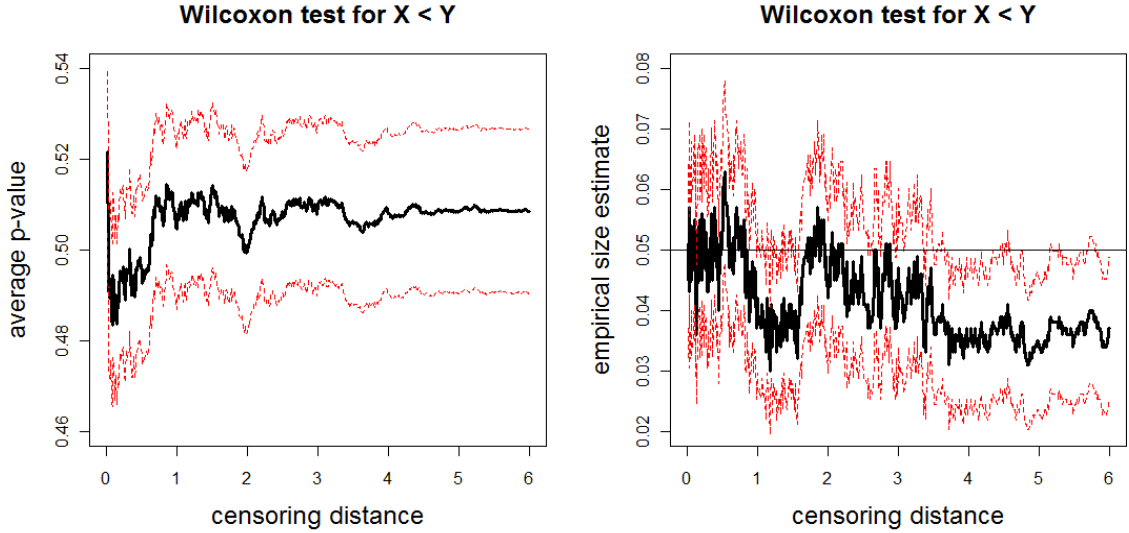

**Figure 7:** The average *p*-values (left) and empirical size estimates (right) versus censoring distance values for pairwise Wilcoxon rank sum test for the left-sided alternative $X < Z$. The average *p*-values and empirical sizes (solid lines) together with the 95% confidence bands (dashed lines) are based on 1000 Monte Carlo replications under the null hypothesis of distributional equality of censored $\mathcal{X}$ and $\mathcal{Z}$ sets that are generated as in Section 2.4.2. Horizontal line in the right plot is at 0.05.

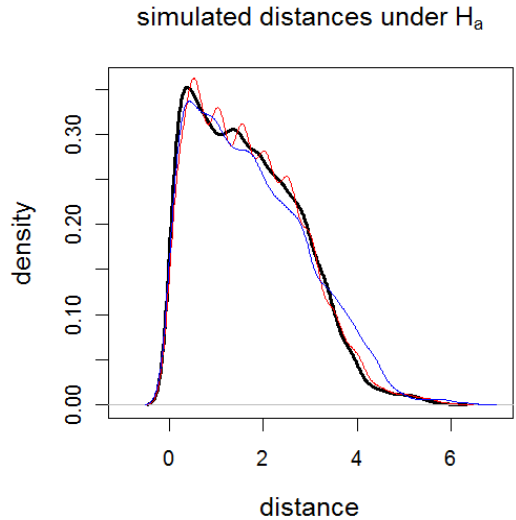

**Figure 8:** Plots of the kernel density estimates of the Monte Carlo distances under the alternative $H_a : r_x = 1.0, r_y = 1.2, r_z = 1.0$ and $\eta_x = 0, \eta_y = 0, \eta_z = 50$. Thick solid black line is for sample $\mathcal{X}$, thin solid red line is for sample $\mathcal{Y}$, and thin solid blue line is for sample $\mathcal{Z}$.

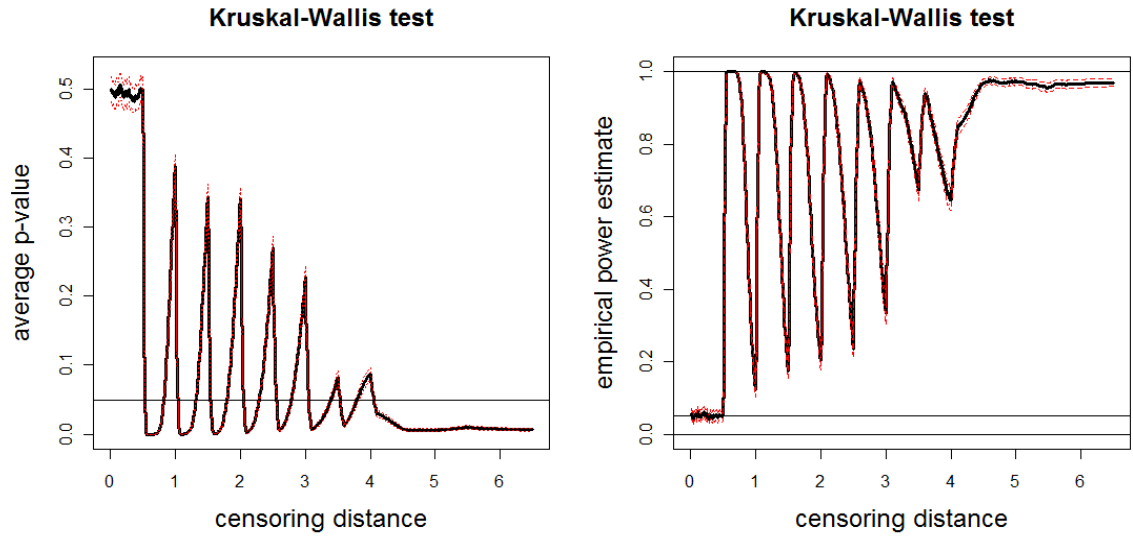

**Figure 9:** Plotted in the left is the average *p*-values ( solid line) and in the right is the empirical power estimates (solid line) versus censoring distance values for multi-group K-W test together with the 95% confidence bands (dashed lines) based on 10000 Monte Carlo replications of censored $\mathcal{X}$, $\mathcal{Y}$, and $\mathcal{Z}$ sets that are generated under the alternative hypothesis (12). Horizontal lines are at 0.05 and 0.95.

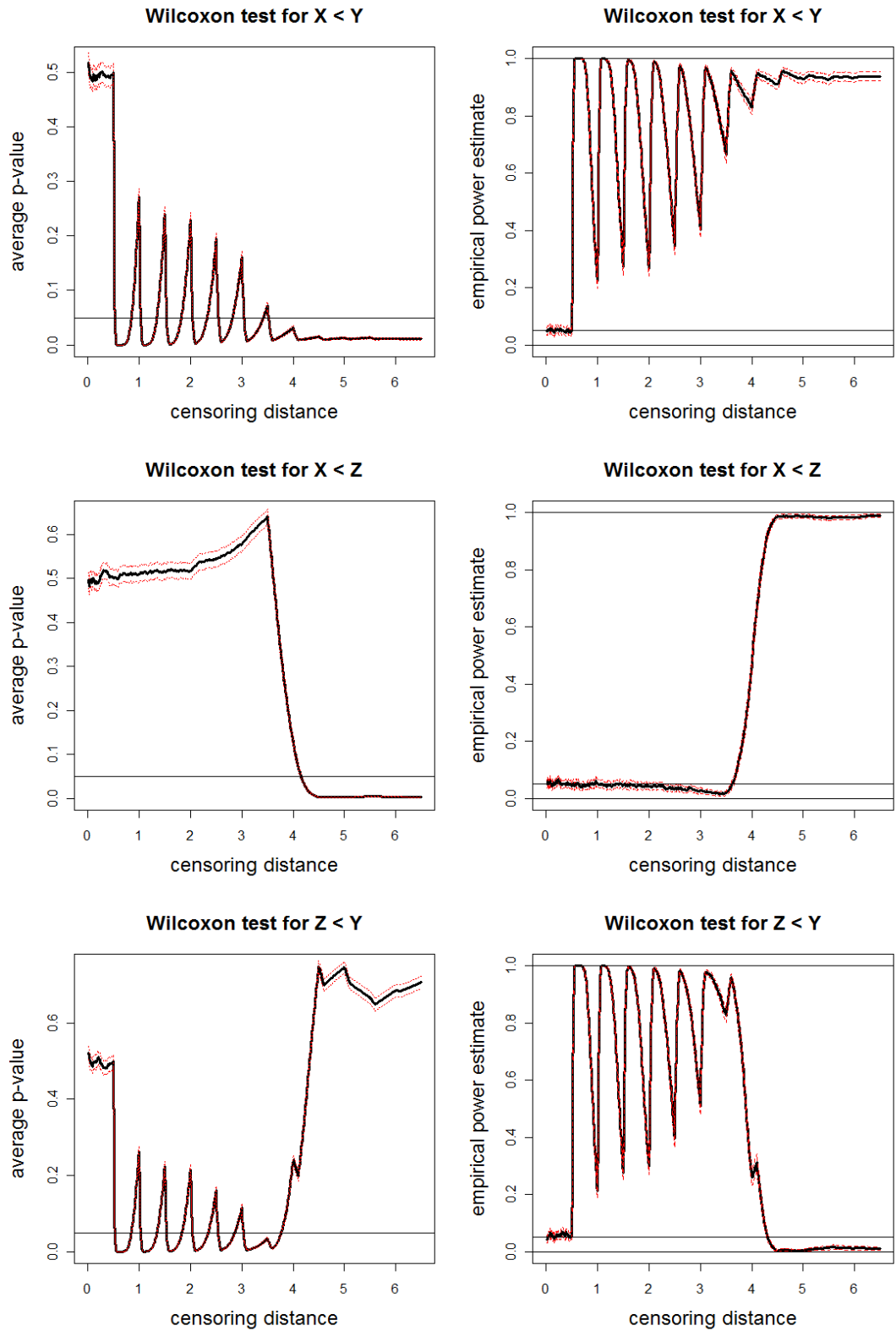

**Figure 10:** Plotted in the left are the average *p*-values (solid lines) and in the right are empirical power estimates (solid lines) versus censoring distance values for Wilcoxon rank sum test for the left-sided alternatives $X < Y$ (top), $X < Z$ (middle), $Z < Y$ (bottom) together with the 95% confidence bands (dashed lines) as in Figure 9. Horizontal lines are at 0.05 and 0.95.